\documentclass[10pt]{iopart} 

\RequirePackage{graphicx}
\usepackage{xcolor}
\RequirePackage{hyperref}
\usepackage{cite}
\usepackage{amssymb}
\usepackage{float}
\usepackage{enumitem}
\usepackage[normalem]{ulem}

\expandafter\let\csname equation*\endcsname\relax

\expandafter\let\csname endequation*\endcsname\relax

\usepackage{amsmath}

\newcommand{\ket}[1]{\left|#1\right\rangle}

\newcommand{\bra}[1]{\left\langle #1\right|}
\newcommand{\bracket}[2]{\left\langle #1|#2\right\rangle}

\newcommand{\be}{\begin{equation}}
\newcommand{\ee}{\end{equation}}
\newcommand{\bea}{\begin{eqnarray}}
\newcommand{\eea}{\end{eqnarray}}

\begin{document}

\title{Modelling assisted tunnelling on the Bloch sphere using the \textit{Quantum Composer}}

\author{Jonas Bley$^1$, Vieri Mattei$^2$, Simon Goorney$^3$, Jacob Sherson$^3$, Stefan Heusler $^{4,*,\dagger}$}

\address{$^1$ Rheinland-Pfälzische Technische Universität Kaiserslautern-Landau, Germany}
\address{$^2$ Purdue Department of Physics and Astronomy, West Lafayette, Indiana, US}
\address{$^3$ Niels Bohr Institute, University of Copenhagen, Denmark}
\address{$^4$ Institut für Didaktik der Physik, Westfälische-Wilhems-Universität Münster, Germany}
\address{$^*$ Correspondence: stefanheusler@uni-muenster.de}
\address{$^\dagger$ Current address: Institut für Didaktik der Physik, Wilhelm-Klemm-Str. 10, 48149 Münster, Germany}

\begin{abstract}
    The Bloch sphere representation is a geometric model for all possible quantum states of a two-level system that can be used to describe the time dynamics of a qubit. As explicit application, we consider the time dynamics of a particle in a double-well potential. In particular, we adopt a recent method for off-resonant excitations, the so-called SUPER principle (Swing-UP of the quantum emitter population) driven by periodic electromagnetic fields, to the context of quantum tunnelling. We show that the tunnelling probability can be enhanced significantly  when an appropriate oscillation of the potential height is introduced. Driven by a collaborative approach we call \textit{educator-developer dialogue}, an updated version of the software Quantum Composer is presented. For educational purposes, we map the two lowest energy states of the 1D-Schr\"odinger equation to the Bloch sphere representation, leading to a rather clear and intuitive physical picture for the pertinent time dynamics.
\end{abstract}


\section{Introduction}

With the advent of quantum technologies, the importance of two-level systems, also denoted as \textit{qubits}, is increasing not only from a technological, but also from a didactical perspective. In many standard textbooks \cite{Sakurai}, the topic is addressed in a rather formal manner. Generally speaking,  multiple representations are an important tool in science education \cite{Gilbert}. This holds true in particular in quantum physics, as the basic concepts are rather abstract and complex. Using the software tool Quantum Composer \cite{R2}, the mathematical formalism comes to life by interactive graphical visualisations. 

In this contribution, as an illustration, we discuss tunnelling in a double well as an explicit example. In particular, using an update of the Quantum Composer, we apply the so-called Bloch-sphere representation to the double-well system. The (single-particle) quantum double-well system is of great importance to many areas of quantum physics as it enables the implementation of a two-level quantum system, i.e. a qubit. Tunnelling in the double-well potential is extensively explored for the symmetric case \cite{Jelic_2012} and the asymmetric (tilted) case \cite{Song_2008}. Periodically driven potentials as examined in this paper are usually described using the Floquet formalism \cite{GH,PhysRevLett.98.263601,PhysRevLett.100.190405,Bukov_2014}.

The Bloch-sphere representation is of particular importance, as it can be applied to all two-level systems, independent of the particular physical interpretation. A geometric model for the time dynamics of qubits on the Bloch sphere is -- for time-independent Hamiltonians -- well known: Just choose a rotation axis ${\vec n}$, which amounts to representing the $2\times 2$ Hamiltonian as $H = \frac{\vartheta}{2} {\vec \sigma}\cdot {\vec n}$. Then, the time dynamics of any initial state $ |\psi \rangle = \alpha |0 \rangle + \beta |1 \rangle$ on the Bloch sphere translates to a rotation of this state around the axis ${\vec n}$ on the Bloch sphere. There are two eigenstates, namely those parallel/anti-parallel to ${\vec n}$. This geometric picture is universal in the sense that it is independent of the explicit physical interpretation of the degrees of freedom described by the Hamiltonian.

A relevant concept to the time dynamics of two-level systems are so-called Rabi oscillations \cite{PhysRev.51.652}. These oscillations describe the behaviour of a two-level quantum systems with an external periodic driving force. This time-dependent driving force leads to an effective coupling between the two states. If the driving force is in resonance with the two-level system, the Rabi oscillations are full, i.e. the particle will oscillate between the two levels completely. Otherwise, the particle will only partially reach the other level. We emphasise that throughout this paper, the term "Rabi oscillations" is used metaphorically as we don't explore systems with time-dependent external driving. 
Rather, in the double well potential, a time-\textit{independent} coupling between the states localised in the left and the right well is present due to the tunnelling, which can only be changed by changing the shape of the potential. Due to the formal equivalence of the time-indepedent and the time-dependent situation, we adopt the notion of "Rabi oscillations" also to the case of tunnelling.

Multiple representations introduced for Rabi oscillations can be useful for undergraduate courses on quantum physics. We provide explicit animations and quantum composer files for full Rabi oscillation (Level 1) and partial Rabi oscillation (Level 2). Next, for graduate level and beyond, we consider a time dependent double-well potential (Level 3). 

This paper is organised as follows:  In Sect. \ref{QC}, the Quantum Composer by Quatomic is introduced.
In Sect. \ref{RO}, we review the well-known general theory for Rabi oscillations of  two-level systems. We discuss the usual co-moving frame approach and contrast it with the situation of time-independent Hamiltonians, where this transformation is not needed. In Sect. \ref{TDW}, the tilted double-well potential is introduced as an explicit model system for a time independent Hamiltonian. We compare analytic results for the degenerate case with numerics from the Quantum Composer (Level 1). Full Rabi oscillations, that is, full oscillation between the left and the right minimum of the double-well potential emerge in this case of a symmetrical double well. In the tilted potential, we have partial oscillation between the left and the right minimum of the double-well.  In Sect. \ref{SUP}, we introduce the so-called SUPER principle on the Bloch-sphere, which generalises the simple model for the time dynamics of a qubit to a situation where the rotation axis ${\vec n}(t)$ on the Bloch sphere itself is a certain function of time, allowing for the transition from a given initial state to a given final state with almost 100\% probability by adjusting ${\vec n}(t)$ accordingly. In Sect. \ref{SUP}, we apply the SUPER principle, which has been introduced in the context of off-resonant excitations \cite{Rei2021}, \cite{Rei2022} to the situation of quantum tunnelling in the tilted double well as model system, which is the main result of the present contribution (Level 3). Note that in contrast to \cite{Rei2021}, we do not use the co-moving frame ansatz. In Sect. \ref{OUT}, we discuss further extentions and applications of this work, and its possible didactical merits.

\section{The Quantum Composer}\label{QC}

\begin{figure}[htb]
    \centering
    \includegraphics[width=\columnwidth]{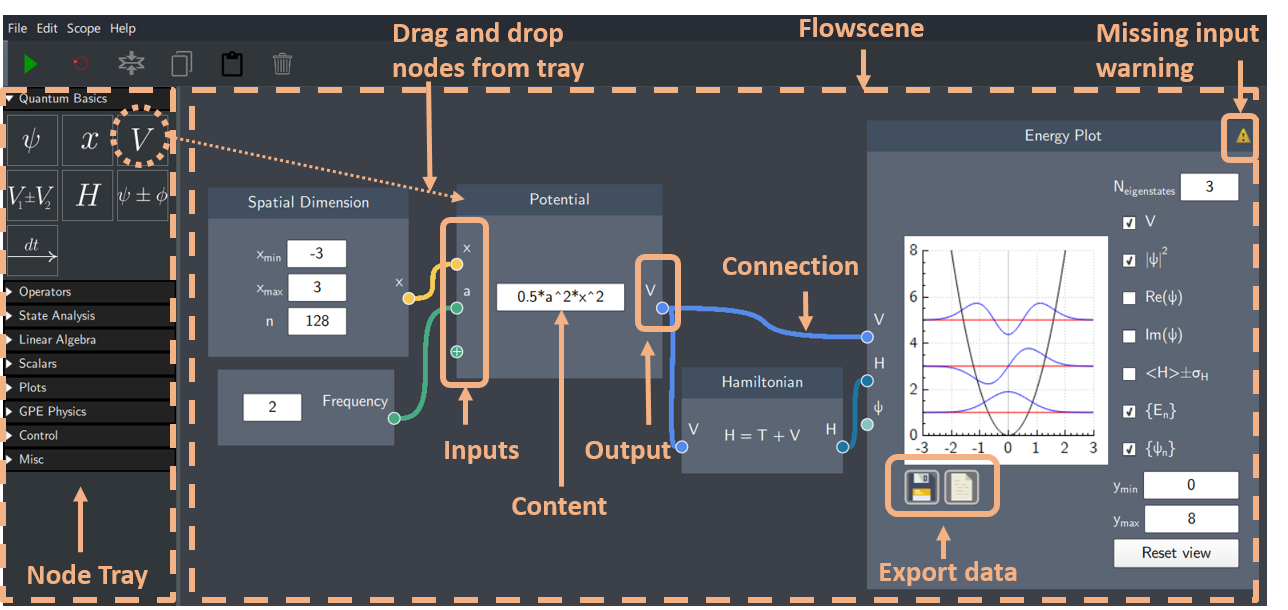}
    \caption{An illustration of the drag-and-drop, node-based, graphical user interface of Quantum Composer.}
    \label{fig:Interface}
\end{figure}

In this work, we will use the digital tool \textit{Quantum Composer} for numerical simulation of the mechanics of a tilted double  well potential. Composer was developed in the group of Prof. Sherson at Aarhus University \cite{quatomic}. Unlike many other fixed-interface applet approaches \cite{kohnle2012new,mckagan2008developing,uni-heidelberg}, which are limited to showcasing particular sets of phenomena, Composer  provides an intuitive and user-friendly re-programmable, node-based graphical interface, as shown in Fig. \ref{fig:Interface}. These nodes can be connected to construct quantum systems which are simulated by the underlying QEngine, a C++ library for the simulation and control of one-dimensional quantum systems \cite{sorensen2019qengine}. These systems under investigation can be generated completely from scratch, or take as a basis pre-defined scenarios loaded into the program using .flow files. This flexibility makes the tool suitable for both education (use by teaching staff and students), and for research purposes \cite{R1,R2,k}, with a low barrier to entry due to the drag-and-drop nature of the programming. The Quantum Composer can be downloaded for free for most desktop platforms \cite{quatomic}.

Due to its hands-on re-programmability, Quantum Composer is uniquely suited to inquiry-based learning (IBL)\cite{a}, in which students interact with didactic material through active discovery, questioning, evaluation, and sensemaking. When used for teaching quantum mechanics, IBL has shown potential for improving understanding and reducing overconfidence \cite{b, c}. Despite this, there are relatively few examples of publicly available resources for inquiry-based quantum mechanics teaching. 

There are many simulation and visualisation tools \cite{d} already available that can be invaluable in promoting understanding, interest, and intuition for quantum concepts \cite{e}. However, those which are most widely used for teaching, such as QuVis \cite{kohnle2012new} from and QuILTs \cite{g} offer a fixed set of scenarios covering different phenomena and allow students to vary parameters. While these are effective to enable educators to set up guided and partially inquiry-based teaching situations, they do not allow for truly open and exploratory inquiry, in which the system is entirely set up from scratch by the participants. Thus, such tools may miss out on some of the benefits associated with more open-ended IBL in understanding and promoting scientific thinking \cite{h}. Composer, however, being entirely reprogrammable, allows for teaching material to be developed which can be anywhere on the spectrum of fully guided to completely open-ended and exploratory. An example IBL approach was previously explored using Quantum Composer to devise student exercises based on the Predict, Explore and Revise approach \cite{Z}.

For researchers, Composer may be useful as a relatively quick and simple means to set up, visualise, and solve analytically intractable quantum systems. The drag-and-drop interface and parameters which can be adjusted on the fly makes it a valuable alternative to more conventional approaches taken by researchers, such as the use of MATLAB or Python code which can be less intuitive to use than the node-based UI of Composer \cite{R1, R2}. The tool has previously been used to investigate the dynamics of atoms in atomic lattices \cite{i} and to solve quantum optimal control problems \cite{sorensen2019qengine, k}.

In the context of this work, however, we wish to highlight another aspect of the Composer re-programmability: the ability of teachers and instructors to seamlessly create custom demonstrators to suit the particular student needs. Here, instructors in advance prepare the relevant .flow files and either use them as lecture demonstrators or distribute them to the students with suitably highlighted experimentation parameters and visualisations preconfigured. In addition, the Composer design team is continually available to support the addition of new interaction and visualisation features. The current work illustrates exactly such interactive collaboration between two instructors (J.B. and S.H.) with the Composer design team resulting in the addition of new features to Composer.

In addition, the Composer design team is continually available to support the addition of new interaction and visualisation features. The current work illustrates a form of collaboration between educators (J.B, S.H), and the Composer developers (V.M, S.G, and J.S) which we call \textit{educator-developer dialogue}. In this interactive approach to co-design of educational material, the digital tool is able to act as an instrument to respond to community needs. 

There are two possible avenues for this dialogue. The first consists of educators designing material with close input from the developers of the tool. This can be invaluable in identifying and preparing validation criteria, as in reference \cite{R2}. The second mechanism, which is demonstrated in this article, is where developers implement new features responding to the needs of the educators, in this case the Bloch-sphere representation. 
Here we demonstrate the value of Composer by extending the body of existing educational material, using the example of Rabi oscillations in the double well potential. Quantum Composer files are provided for three different levels: full Rabi oscillations (Sect. \ref{sec:resrabi}), partial Rabi oscillations (Sect. \ref{sec:nonresrabi}) and Swing Up with the SUPER principle (Sect. \ref{SUP}). The corresponding files can be downloaded from \cite{Animations}. By combining the Bloch sphere representation with the representation of wave functions in the double-well, Composer helps us to develop an intuitive physical picture of the time dynamics in the time-dependent double well using multiple representations. This is the first application of an update of the Quantum Composer incorporating the Bloch sphere representation, driven by an educator-developer dialogue.

\section{A brief review of the general theory of Rabi oscillations}
\label{RO}

In this section, we briefly review the general theory of Rabi oscillations, which holds for many different applications. Crucially, we distinguish between two-level systems driven by periodic fields, and those that are not driven by periodic fields. 

\subsection{Two-level systems driven by periodic fields}
\label{RoTAA}

We start our discussion with the usual textbook example of a two-level systems driven by a periodic field.  The unperturbed states are defined as $| g \rangle $ (ground state) and $| e \rangle$ (excited state) at energies $E_g$, $E_e$. Let $\Delta e^{i \Omega t}$ be the interaction between the driving field and the two-level system that introduces a coupling between the two eigenstates, where $\Omega$ is the driving frequency of the periodic field and $A$ the (complex) amplitude. This system is described by the following Hamiltonian
\begin{eqnarray}
\label{EZH}
H_1(t) = E_g | g \rangle \langle g | + E_e | e \rangle \langle e | + \frac{1}{2}A e^{i \Omega t}| g \rangle \langle e | + \frac{1}{2} A^* e^{-i \Omega t}| e \rangle \langle g |.
\end{eqnarray}
In matrix notation, we obtain (with mean energy $\bar{E}= (E_e+E_g)/2$ and energy gap $\eta \equiv E_e-E_g$ between excited and ground state)
\begin{eqnarray}
\label{OrHa}
H_1(t) =
  \left[ {\begin{array}{cc}
   E_g &     \frac{1}{2} A e^{i \Omega t}  \\
 \frac{1}{2}  A^* e^{- i \Omega t} &      E_e  \\
  \end{array} } \right] =  \bar{E}\cdot I +   \frac{1}{2} \left[ {\begin{array}{cc}
  \eta &     A e^{i \Omega t}  \\
  A^* e^{- i \Omega t} &    - \eta \\
  \end{array} } \right]
\end{eqnarray}
where $I$ is the unit matrix. The Schr\"odinger equation describing the dynamics of the system is given by $i \hbar \partial_t |\psi \rangle = H_1(t) | \psi \rangle$. An elegant way to compute the time dynamics of the state $| \psi(t) \rangle = z_g(t) |g \rangle + z_e(t) |e \rangle $ is the transformation to the so-called co-moving frame: Using the unitary operation $C(t) = \exp{[- i (\Omega t/2) \sigma_3]}$, where $\sigma_3$ is the third Pauli matrix, we define by $|\phi \rangle \equiv C(t) |\psi \rangle$ the state in the co-moving frame. Within the co-moving frame, the time-dependence of the state $| \phi (t) \rangle$ is then given by
\bea
i \hbar \partial_t |\phi \rangle = (C(t) H_1(t) C^{\dagger}(t) - i \hbar C(t) \partial_t C^{\dagger}(t)) |\phi \rangle \equiv H_{\rm CM} | \phi \rangle 
\eea
with
\bea\label{CoM}
H_{\rm CM} =
 \bar{E} +   \frac{1}{2} \left[ {\begin{array}{cc}
  \eta  - \hbar \Omega &     A  \\
  A^*  &    - \eta  + \hbar \Omega \\
  \end{array} } \right].
\eea
In the co-moving frame, the Hamiltonian $H_{\rm CM}$ becomes time-independent, while the diagonal terms $\pm \eta$ are shifted to $\pm (\eta - \hbar \Omega)$. That is, the time dependence $\Delta e^{i \Omega t}$ of the interaction is transformed to a change in the energy shift in the co-moving frame. Resonance emerges for $\eta \equiv E_e - E_g = \hbar \Omega$ which can then be interpreted as absorption or emission of a photon in the transition between the states $|e \rangle \leftrightarrow |g \rangle$.

\subsection{Two-level systems without external driving by periodic fields}

In contrast to the textbook example for Rabi oscillations induced by external periodic fields, in this contribution, we consider two-level systems without external driving. In other words, the Hamiltonian is time-independent from the very beginning, and no transformation to the co-moving frame is necessary in order to achieve this. Due to the formal equivalence between (\ref{OrHa}) and (\ref{CoM}), all mathematical results can be adopted, while the interpretation changes significantly. Note that in our approach to tunnelling, we therefore do \textit{not} work in the co-moving frame.

Consider a time-independent two-level system with states $| L \rangle $ and $| R \rangle$ at energies $E_L$, $E_R$ with mean energy $\bar{E}$ and energy difference $\delta = E_L-E_R$. Let $\Delta$ be the (real) interaction between the two states (which will later on be interpreted as transition probability between the left and right potential minimum, see Eq. (\ref{coup})). This system is described by the following time-independent Hamiltonian

\begin{eqnarray}
\label{H0}
H_0 &=&
  \bar{E}  + 
\frac{1}{2}  \left[ {\begin{array}{cc}
  \delta &      \Delta   \\
   \Delta &      -\delta  \\
  \end{array} } \right] \nonumber \\
  &=&
   \bar{E}  + 
\frac{\sqrt{\delta^2 + \Delta^2}}{2}  \left[ {\begin{array}{cc}
  \frac{\delta}{\sqrt{\delta^2 + \Delta^2}} &        \frac{\Delta}{\sqrt{\delta^2 + \Delta^2}}  \\
  \frac{\Delta}{\sqrt{\delta^2 + \Delta^2}} &      -  \frac{\delta}{\sqrt{\delta^2 + \Delta^2}} \\
  \end{array} } \right] 
  \nonumber \\ 
  &\equiv& \bar{E}  + \frac{\hbar\omega_R}{2}
  \left[ {\begin{array}{cc}
  \cos(2 \theta) &      \sin(2 \theta)   \\
   \sin(2 \theta) &      - \cos(2 \theta)  \\
  \end{array} } \right] = \bar{E}  + 
\frac{\hbar\omega_R}{2} \vec{\sigma}\cdot {\vec n}.
\end{eqnarray}
First, we want to exploit the general results following from this Hamiltonian. Later on, we use this Hamiltonian to approximate the two lowest energy states of the double-well potential. 
We introduced the so-called Rabi frequency $\hbar \omega_R = \sqrt{\delta^2 + \Delta^2}$ and the rotation angle $\theta$, which describes the rotation axis of the corresponding time development in the unitary operator 

\begin{eqnarray}
\label{UnOp}
U(t) = \exp[- i H t/\hbar] =\exp[- i {\bar E} t/\hbar - i \frac{\omega_R t}{2 } {\vec \sigma}\cdot{\vec n}].
\end{eqnarray}
$\vec \sigma$ are the three Pauli-matrices. Indeed, on the Bloch sphere, this operator generates a rotation of any given initial state around the rotation axis 

\bea
\label{RotA}
{\vec n} = (\sin(2 \theta), 0, \cos(2 \theta)) \equiv (\frac{\Delta}{\hbar\omega_R}, 0, \frac{\delta}{\hbar\omega_R}).
\eea
On the Bloch sphere, the eigenstates point in the $\pm {\vec n}$ directions. In Hilbert space, these eigenstates are given by
\bea
\label{EST}
|+ {\vec n} \rangle = \left[ {\begin{array}{c}
  \cos(\theta)  \\
   \sin( \theta) \\
  \end{array} } \right], \ \ \ \ \  |- {\vec n} \rangle = \left[ {\begin{array}{c}
  - \sin(\theta)  \\
   \cos( \theta) \\
  \end{array} } \right].
\eea
The eigenvalues are given by $E_{1} = {\bar E} + \frac{1}{2}\hbar\omega_R$ and  $E_{2} = {\bar E} - \frac{1}{2}\hbar\omega_R$, respectively. We find for $|L \rangle$ and $|R \rangle$
 \bea
 \label{Tat}
 |L \rangle &=& \cos(\theta) |+{\vec n}\rangle - \sin(\theta) |-{\vec n}\rangle ,
 \\ \label{Tat2}
  |R \rangle &=& \sin(\theta) |+{\vec n}\rangle + \cos(\theta) |-{\vec n}\rangle.
 \eea

\begin{figure}[htb]
\centering
\includegraphics[width=0.95\columnwidth]{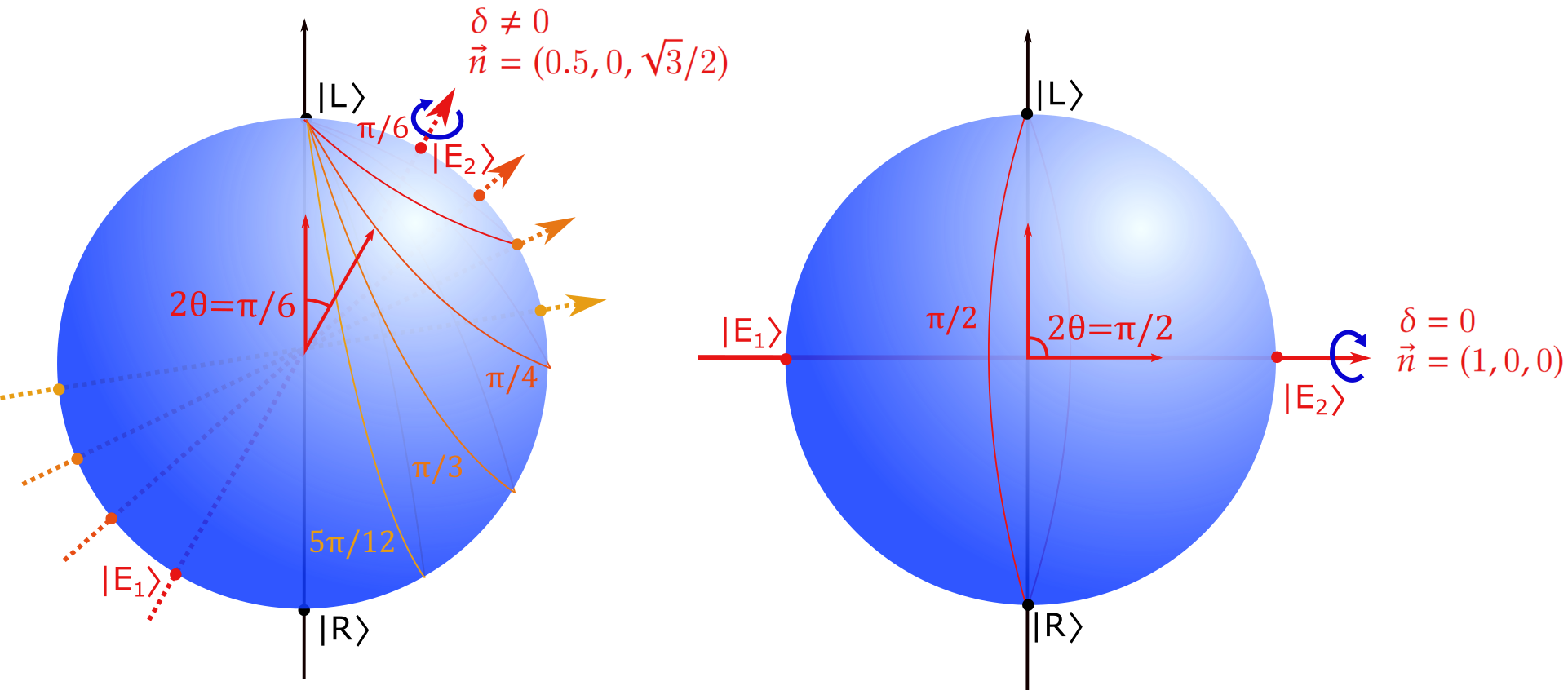}
\caption{Coloured Orbits on the Bloch sphere for different rotation axis angles $2\theta$. The rotation axis $\vec{n}$, defined by the first two energy eigenstates of the system $E_1$ and $E_2$, is shown in red. Left: Partial Rabi oscillations, time development of the initial state $| L \rangle$ for $2 \theta =  \pi/6, \pi/4, \pi/3$ and $5\pi/12$. Right: Full Rabi oscillation, $2 \theta = \pi/2$. In this case, the transition probability to state $| R \rangle$ is $100\%$ after a $\pi-$flip.}
\label{fig:Blochspheres2}
\end{figure}

For a given initial state $| \psi(0) \rangle$, the fidelity under time development $| \psi(t)\rangle = U(t) | \psi(0)\rangle$ is defined as $F(t)=|\langle \psi(0) | \psi(t) \rangle|^2$. For a general superposition state 
\bea
|\psi_0 \rangle = \alpha |+{\vec n} \rangle + \beta |- {\vec n} \rangle
\eea
with $|\alpha|^2+|b|^2 = 1$, the time dynamics is given by (up to a global phase)

\bea \label{eq:timedynamics}
|\psi(t) \rangle &= e^{- \frac{i}{2} \omega_R t \vec{\sigma} \cdot{\vec n}} (\alpha |+{\vec n} \rangle + \beta |- {\vec n} \rangle) \nonumber \\ &=  e^{+ \frac{i}{2} \omega_R t} \alpha |+{\vec n} \rangle + e^{- \frac{i}{2} \omega_R t} \beta |-{\vec n} \rangle
\eea
and the fidelity, i.e. the tunnelling probability, reads 
\bea \label{FIDE}
F &= |\langle \psi(0) | \psi(t) \rangle|^2 = |e^{+ i\frac{1}{2} \omega_R t} \alpha^2 + e^{- i \frac{1}{2}\omega_R t} \beta^2|^2 \\ &= 1 - 4 |\alpha|^2 |\beta|^2 \sin^2(\frac{\omega_R t}{2}).
\eea
For $| \psi(0) \rangle = |L \rangle$, we obtain
\bea
\label{Fide}
F =  1 - \sin^2(\frac{\omega_R t}{2}) \sin^2(2 \theta).
\eea
For $2 \theta = \pi/2$, the Rabi oscillation is maximal. The fidelity is then given by
\bea
\label{FRes}
F = 1 - \sin^2(\frac{\omega_R t}{2}).
\eea

The state $|\psi_0 \rangle = \alpha |+{\vec n} \rangle + \beta |- {\vec n} \rangle$ can be mapped on the Bloch sphere via Hopf mapping (the qubit vector in $\mathbb{C}^2$ is mapped to the sphere $\mathcal{S}^2$, see e.g. \cite{hopf_fib}). Explicitly, 

\bea
(\alpha, \beta) \rightarrow (\rm{Re}(2 \alpha^* \beta), \rm{Im}(2 \alpha^* \beta), |\alpha|^2 - |\beta|^2).
\eea
On the Bloch sphere, there is a simple geometric interpretation for the behaviour of the fidelity. In Fig. \ref{fig:Blochspheres2}, we show the rotation axis ${\vec n}$ of the unitary time development, and the orbits of states at angle $\theta = 0, \pi$ (the eigenstates), and $\theta_k = k \pi/6, \ k = 1, \ldots 5$. 
In analogy to the time dependent system described, for $2 \theta = \pi/2$, $\ket{L}$ and $\ket{R}$ are part of the same orbit. Therefore, the transition between the two states is certain, leading to the fidelity function (\ref{FRes}). We therefore call this \textit{full} Rabi oscillation, in analogy to the time dependent system described.

\section{Tunnelling in the double well}
\label{TDW}

Quantum tunnelling plays a major role in various areas of modern physics. Early applications have been e.g. the calculation of lifetimes for alpha-decay of nuclei, or for nuclear fusion. In recent times, from a technological perspective, tunnelling in solid state physics, and in particular in semiconductor physics and in spintronics \cite{Miao} are of interest both for the first and in the second quantum revolution. Furthermore, model systems such as Bose-Einstein condensates in a double-well potential are part of actual research \cite{AB, DWM}. In the past decades, periodically driven tunnelling systems also have been widely studied \cite{GH}. Since tunnelling is an important topic in physics since the advent of quantum physics, it is not astonishing that also in physics education, quantum tunnelling is part of the standard curriculum not only at university, but also in high school \cite{Stade}. Thus, it is desirable to explore model systems for quantum tunnelling at all levels of education, and also for research.

\subsection{Full Rabi oscillations in the double well (Level 1)}\label{sec:resrabi}
As a model for the tunnelling system, we consider the (tilted) double-well potential 
\bea
\label{DW1}
V(x) = \lambda (x^2-a^2)^2+\mu x. 
\eea
If we consider the system to consist of two subsystems L and R, we find respective ground states $\ket{L}$ and $\ket{R}$. First, we want to find a relation between the parameters $(\delta, \Delta)$ of the general theory of Sect. \ref{RO}, and the parameters $(\lambda, a)$ of $V(x)$. For this purpose, we investigate the degenerate case $\mu=\delta=0$ where the ground state energies $E_L$ and $E_R$ are equal.
\begin{figure}
    \centering
    \includegraphics[width=0.95\columnwidth]{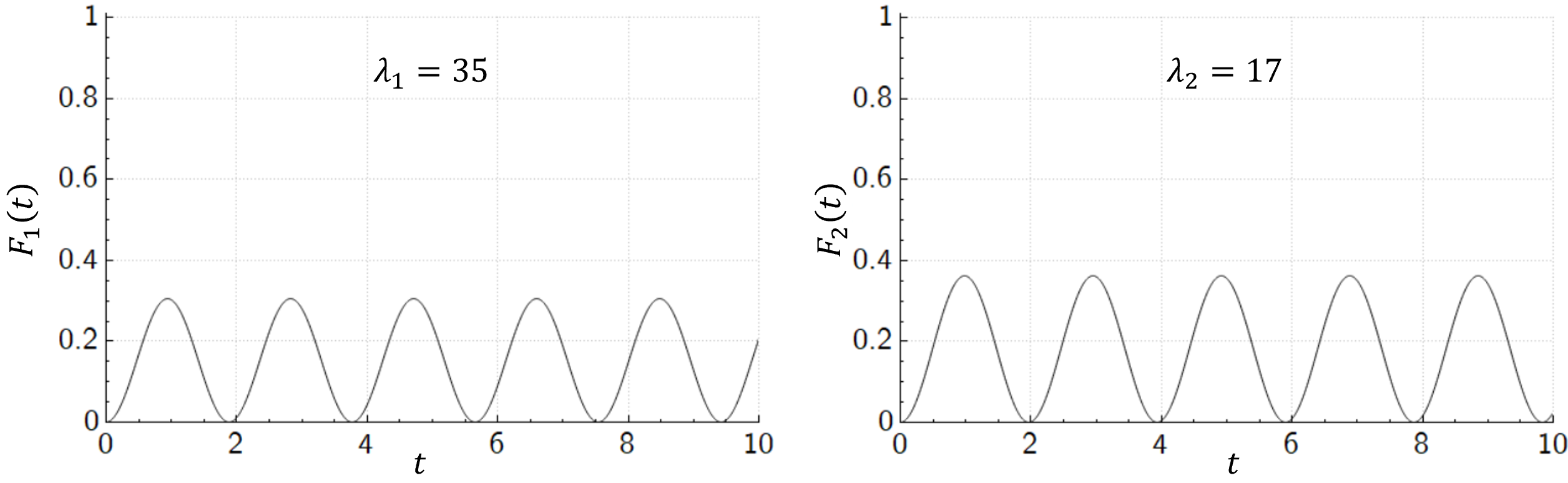}
    \caption{Tunnelling probabilities for two different potentials with $a=0.7$ and $\mu=2.8571$. Left: $V_1$ with $\lambda_1=35$, $1-F_1(t)=1-\left|\bracket{L_1}{\psi_1(t)}\right|^2$ with $\psi_1(0)=\ket{L_1}=0.9575\ket{E_2}+0.2884i\ket{E_1}$. Right: $V_2$ with $\lambda_2=17$,: $1-F_2(t)=1-\left|\bracket{L_2}{\psi_2(t)}\right|^2$ with $\ket{L_2}=\psi_1(0)=0.9483\ket{E_2}+0.3172i\ket{E_1}$. These two values of $\lambda$ will be used to realise assisted tunnelling with the SUPER principle which is shown in Sect. \ref{SUP}. The theoretical prediction for the fidelity $1-F(t) = 4 |\alpha|^2 |\beta|^2 \sin^2(\omega_R t/2)$ coincides with the numerical result and cannot be distinguished within the resolution of these plots.}
    \label{fig:lambda1lambda2}
\end{figure}

We will calculate $\Delta = \Delta(\lambda, a)$ within this model potential. In the degenerate case, we obtain full Rabi oscillation between the state on the left $\ket{L}=1/\sqrt{2}(\ket{E_2}-\ket{E_1})$ and on the right side of the potential $\ket{R}=1/\sqrt{2}(\ket{E_2}+\ket{E_1})$ (see Eq. \ref{Tat} and \ref{Tat2} for $\theta=\pi/4$). The eigenenergies are given as $E_1 = {\bar E} + \sqrt{\delta^2 + \Delta^2} \rightarrow {\bar E} + \Delta$ and $E_2 = {\bar E} - \sqrt{\delta^2 + \Delta^2} \rightarrow {\bar E} - \Delta$. In this case, the coupling $\Delta$ can be computed as tunnelling amplitude either using WKB or the instanton ansatz. In both approaches, we obtain \cite{Rast, Cole2}
\bea
\label{coup}
\frac{1}{2} \Delta (\lambda, a, \mu=0) &= \sqrt{12} \hbar \omega_{H.O.} \sqrt{\frac{S_{\rm inst}}{2 \pi \hbar}} e^{- \frac{S_{\rm inst}}{\hbar}} \nonumber \\ &= \sqrt{\frac{m^2 \omega_{H.O.}^5}{ 2 \pi \lambda}} \exp{[- \frac{1}{\hbar}(\frac{4}{3} \sqrt{2 m \lambda} a^3)]}.
\eea
Here, $x_{\rm inst}$ is the so-called instanton solution, a classical path in imaginary time $\tau$ connecting the left and the right potential minimum.  The instanton action is given by $S[x_{\rm inst}] = \sqrt{2 m} \int \sqrt{V(x)} dx$, leading to
\bea
\label{Cole}
S_{\rm inst.} = \sqrt{2 m} \int_{-a}^{+a} \sqrt{\lambda} |x^2-a^2| d x = \frac{4}{3} \sqrt{ 2 m \lambda} a^3 \equiv \frac{m^2 \omega_{H.O.}^3}{12 \lambda}
\eea
with $\omega_{H.O.} = \sqrt{(8 a^2 \lambda)/m}$. In such a way, we find explicit and analytic expressions for all parameters in the general Hamiltonian (\ref{EZH}) in terms of the model potential (\ref{DW1}) for the degenerate case $\mu = 0$.

\subsection{Partial Rabi oscillations in the tilted double well $\mu\neq0$ (Level 2)}\label{sec:nonresrabi}

As an approximation, we can consider the two wells to be completely separate, i.e. the system to consist of two independent subsystems L and R. Let $E_L$ be the energy of the ground state $\ket{L}$ of the left minimum, and $E_R$ be the energy of the ground state $\ket{R}$ of the right minimum. Concerning these two ground state energies, we may approximate the respective potentials as harmonic oscillators. We find in the vicinity of $x \simeq  a$ with $y \equiv (x - a)$
\bea
\label{VT}
V(y) \simeq \mu (y+a) + (2 a)^2 \lambda y^2  + \ldots 
\eea
In first order approximation, the $\mu$-dependence is irrelevant for the ground state apart from a vertical shift by $-\mu a$ or $\mu a$, respectively, as the term $\mu(x\pm a)$ is odd around $x=\mp a$, and the harmonic oscillator ground state is even around its origin. Thus, we can compare the ground state energies with the usual ground state energy of the harmonic oscillator. We find $E_L = \frac{1}{2} \hbar \omega_{H.O.} - \mu a$ and $E_R = \frac{1}{2} \hbar \omega_{H.O.} + \mu a$ with $\hbar \omega_{H.O.} = \sqrt{(8 a^2 \lambda/m)}$. Thus, we find in leading order $\delta = (E_L - E_R)  \simeq - 2 \mu a$. While it is possible to refine the (rather bad) approximations for $E_{L/R}$ using asymptotic series \cite{Rast}, we refrain from doing so here as our main concern is the introduction of assisted tunnelling using the Quantum Composer.

We will discuss to what extent we can verify the theory of Rabi oscillations using the Quantum Composer. This is the second level of studying Rabi oscillations in the double-well potential with the Quantum Composer. Natural units ($\hbar=1$ and $m=1$) are used throughout the rest of this paper. Since the eigenvalues for the energy are $E_1 = {\bar E} + \frac{1}{2} \sqrt{\Delta^2 + \delta ^2}$ and $E_2 = {\bar E} - \frac{1}{2} \sqrt{\Delta^2 + \delta ^2}$, we can determine the Rabi frequency $\omega_R=\sqrt{\Delta^2+\delta^2}$ from $E_1 - E_2 = \omega_R$ and the mean energy by $E_1 + E_2 = 2 {\bar E}$, obtaining $E_1$ and $E_2$ numerically using the Quantum Composer. In Fig. \ref{fig:lambda1lambda2}, we compare the theoretical expression for the fidelity $1-F(t) = 4 |\alpha|^2 |\beta|^2 \sin^2(\omega_R t/2)$ (\ref{FRes}) with the results from the quantum composer for two different potentials ($\lambda_1= 35, \lambda_2=17$). We find excellent agreement.

In the tilted potential, the instanton calculations that were shown in Sect. \ref{sec:resrabi} for full Rabi oscillations are still possible, but much more involved \cite{Rast}. We will not embark into these troubled waters of theory not to loose track to our main purpose, the introduction of assisted tunnelling. However, it would be interesting for future work to link also these analytic results with numerical results obtained with the Quantum Composer.

To conclude, Rabi oscillations can be studied in an elegant manner with the Quantum Composer for the tilted double well potential and can give valuable insights into some important aspects of the underlying theory.

\begin{figure}[htb]
\centering
\includegraphics[width=0.95\columnwidth]{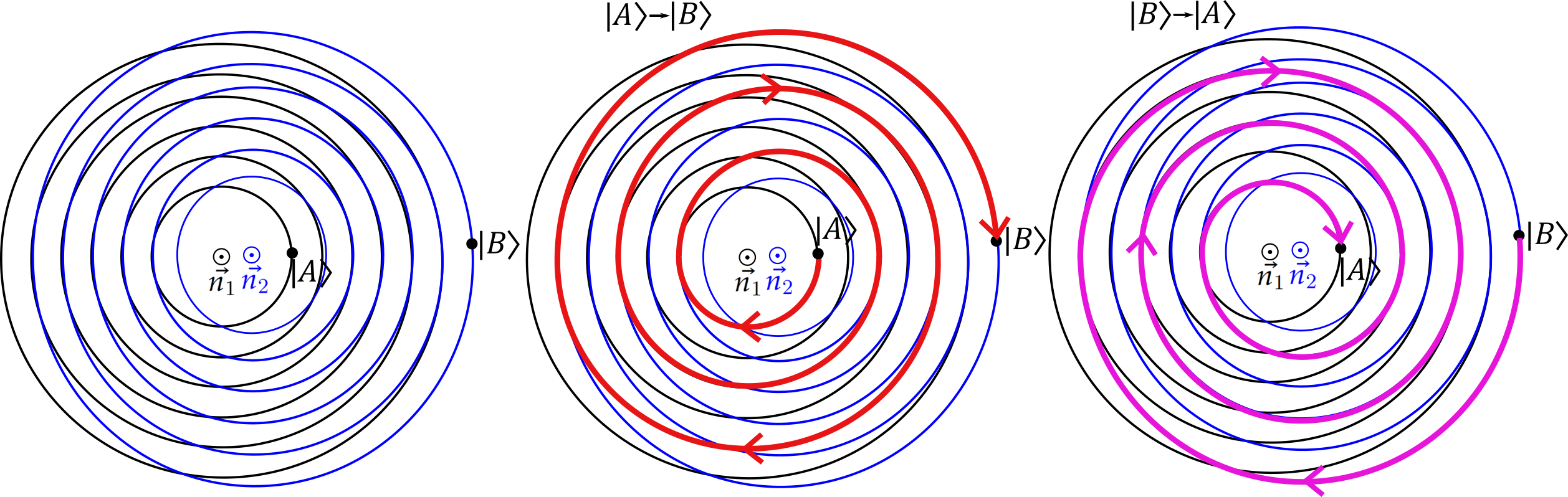}
\caption{The Swing Up principle illustrated in 2D. Two different potentials $V_1$ and $V_2$ have two different rotation axis $\vec{n}_1$ and $\vec{n}_2$ pointing towards the reader. By swapping between them, we can go from initial state $\ket{A}$ to state $\ket{B}$ following the red path, or we can go from $\ket{B}$ to $\ket{A}$ following the pink path. When falling out of the swinging rhythm, there is the danger to 'loose track'. Once the state is at $\ket{B}$, if the Swing Up process is not stopped it will follow the pink path and start decaying back to $\ket{A}$. Note that in a full Swing Up, a state viewed in 2D would evolve along the red path and then "back" along the pink path to end in the desired state, orthogonal to the initial state.}
\label{fig:2Dswingup}
\end{figure}

\section{Application of the SUPER-principle to assisted tunnelling (Level 3)}
\label{SUP}

The SUPER principle was originally introduced in \cite{Rei2021}, while first experimental realisations have been reported in \cite{Rei2022} for two level-systems driven by periodic fields. Note that the rotating frame transformation must be applied in this case. 

In the rotating frame, the time-independent Hamiltonian (5) arises. As the tunnelling system is of the type of equation (5) from the very beginning, no transformation to the rotating frame is necessary. This necessity is the only difference between the original SUPER principle and the ideas applied here. This allows for the application to assisted tunnelling, as we will show in what follows.

The key idea of the SUPER principle is the following: For given rotation axis ${\vec n}$, the orbit of all initial states under the action of (\ref{UnOp}) form a "ring system" as shown in Fig. \ref{fig:2Dswingup}. If the rotation axis ${\vec n}(t)$ oscillates between \textit{two different}, but near positions, the time development of an initial state switches between these two "ring systems". 

\begin{figure}[htb]
\centering
\includegraphics[width=0.85\columnwidth]{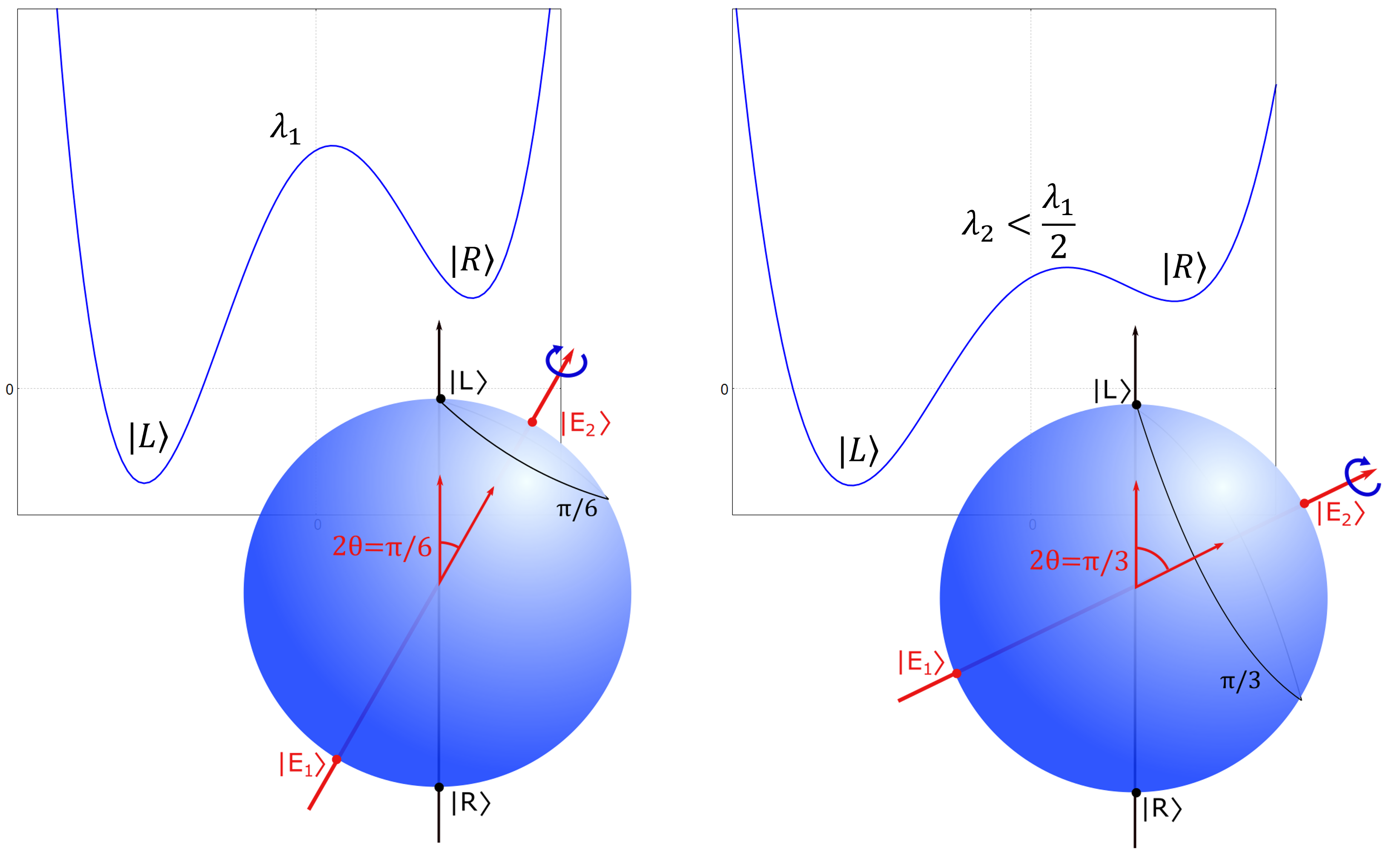}
\caption{Relation between potential barrier height and the corresponding time dynamics on the Bloch sphere. By switching between these two potentials, on the Bloch-sphere, the SWING up mechanism can be realised as shown in Fig. \ref{fig:2Dswingup} } 
\label{fig:Pot-TD}
\end{figure}

\begin{figure}
    \centering
    \includegraphics[width=0.95\columnwidth]{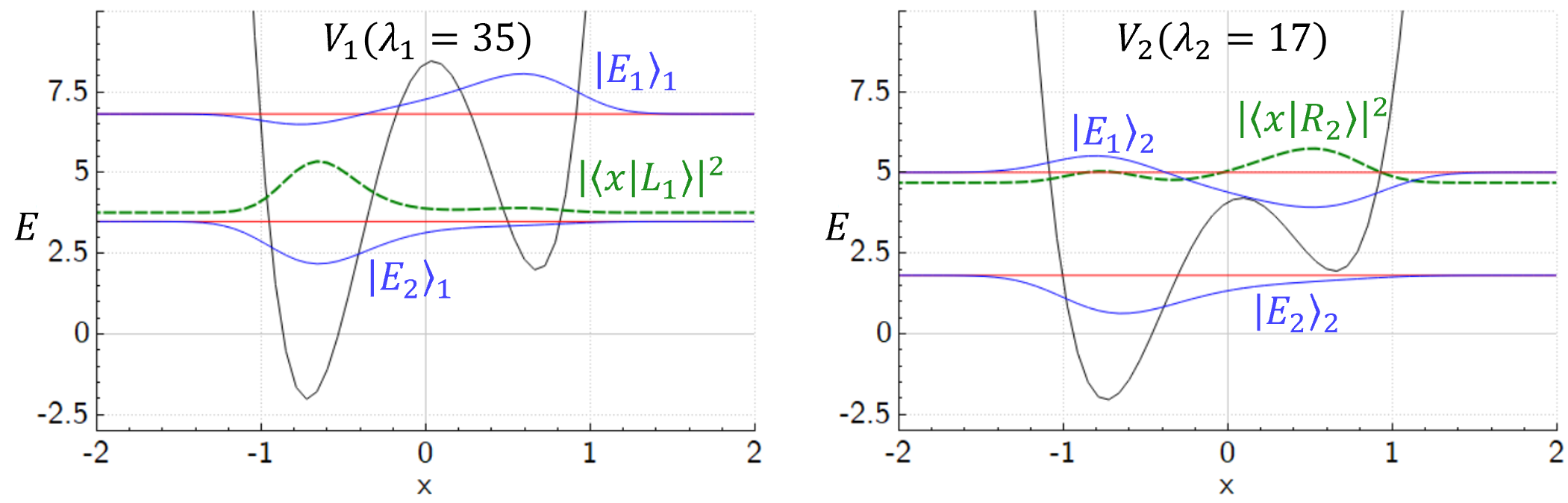}
    \caption{Two potentials $V_1$ with $(\lambda_1=35,a_1=0.7,\mu_2=2.8571)$ and $V_2$ with $(\lambda_2=17,a_2=0.7,\mu_2=2.8571)$ with the probability density function of the initial state $\ket{L_1}=0.9575\ket{E_2}_1+0.2884i\ket{E_1}_1$ and the desired state $\ket{R_2}=0.2884i\ket{E_2}_2+0.9575\ket{E_1}_2$. The eigenstates $\ket{E_{2/1}}_1$ of $V_1$ and $\ket{E_{2/1}}_2$ of $V_2$ are shown in blue, with eigenenergies in red. The corresponding Bloch-sphere representation is shown schematically in Fig. \ref{fig:Pot-TD}}
    \label{fig:SwingUpPotentials}
\end{figure}

In \cite{Animations}, we show an animation of the time evolution of the state $| L  \rangle$ to the state $| R  \rangle$ based on the super principle, applied to the situation shown in Fig. \ref{fig:Pot-TD}. Indeed, as we show in this section, if the oscillation period is chosen nearby the average Rabi frequency of the two different unitary operators (\ref{UnOp}) with rotation axis ${\vec n}_1$ and ${\vec n}_2$, a "Swing Up" leads to an enhancement of the Rabi amplitude even in the case of a tilted well. From our point of view, this result can be seen as analogous to \cite{Rei2021} in the case of Rabi oscillations with an external, time-dependent driving field, where it was shown that the modulation frequency has to be chosen close to the mean Rabi frequency of the two detunings.
In order to apply the SUPER principle to the particular situation of tunnelling in a double-well, we consider a situation where the potential height $V(0)= \lambda a^4$  for constant distance $a$ is periodically shifted from a 'high' value to a 'low' value by changing $\lambda$. The potentials used, and the initial and desired states are shown in Fig. \ref{fig:SwingUpPotentials}. We consider the situation where we start in state $| L_1 \rangle$ with $V_1$ or $| L_2 \rangle$ with $V_2$. For large tilting parameter $\mu$, the maximum fidelities $F_1=|\langle L_1 | R_1 \rangle|^2$ and $F_1=|\langle L_2 | R_2 \rangle|^2$ are rather low (see Fig. \ref{fig:lambda1lambda2} with $P_1=0.304$ and $P_2=0.362$, respectively).

\begin{figure}
    \centering
    \includegraphics[width=\columnwidth]{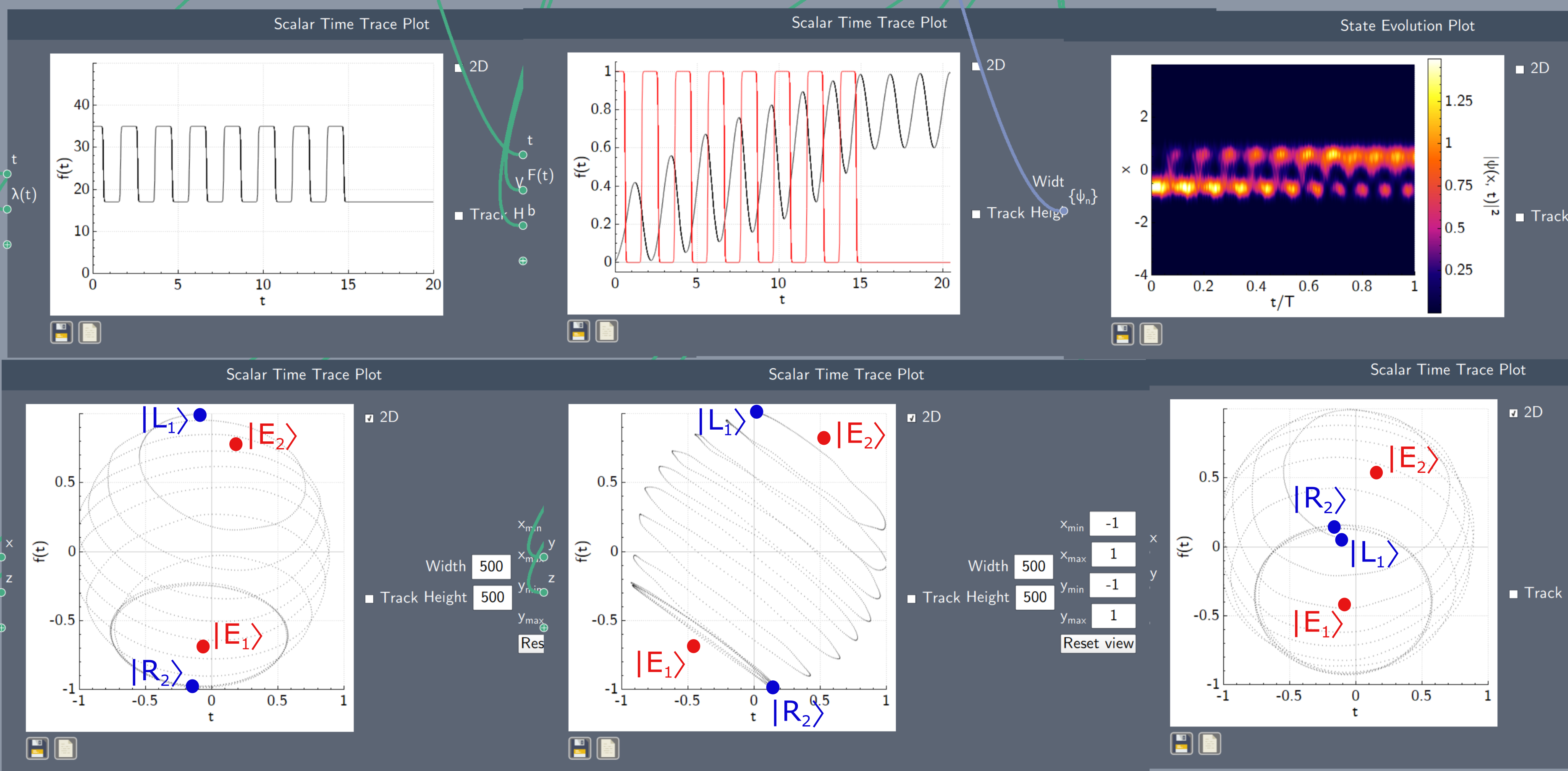}
    \caption{Assisted tunnelling by oscillation between $V_1$ with $\lambda_1=35$ and $V_2$ with $\lambda_2=17$  at potential swap time (the time one potential is active before swapping to the next) $T/2=1.01$. Top Left: Potential barrier height $\lambda(t)$. At 15 time units, the potential was held constant to trap the state at the top. Top Middle: Tunnelling probability $F(t)=\left|\bracket{\psi(t)}{R_2}\right|^2$ with $\psi(0)=\ket{L_1}$.  Top Right: State evolution plot. Bottom: Look at the Bloch Sphere with initial state $\ket{L_1}$, desired state $\ket{R_2}$ and the eigenstates $\ket{E_2}$, $\ket{E_1}$ "from above" (Left), "from the front" (Middle, comparable to Fig. \ref{fig:Blochspheres2}) and "from the side" (Right).
    By periodic oscillation between $\lambda_1=35$ and $\lambda_2=17$ and trapping the state at $t=15$ at $\lambda_2=17$, a maximum fidelity of over 0.994 was achieved at $t=20.5$.}
    \label{fig:assistedTunnelling1}
\end{figure}

Next, we modify the system and allow for an oscillation between $V_1$ or $V_2$, where the potential oscillation frequency is about the mean Rabi frequency. We introduce a smoothstep function to model the transition between the potentials. After some fine tuning that is described in \ref{SUP2}, the maximum Fidelity was increased drastically to $P=0.994$. The result is shown in Fig. \ref{fig:assistedTunnelling1}, confirming the Swing up mechanism, leading to assisted tunnelling.

Besides the SUPER principle, other approaches to enhance the tunnelling rate are known. Here, the notion of parametric resonance is used in a variety of different situations. Our approach differs from parametric resonance as introduced in \cite{Peano}, as we consider a \textit{tilted} potential, and change the coupling between the two wells in an appropriate way by changing the barrier height. We do not use the rotating wave approximation.
In \cite{Golo_2005}, the amplitude of an external driving field is changed to enhance the tunnelling rate. In both cases, parametric resonance works with a weakly nonlinear, parametrically modulated oscillator described in the rotating wave approximation with tilting $\mu = 0$.

Other approaches to enhance tunnelling -- also denoted parametric resonance -- have been proposed in \cite{Salasnich_2002} and \cite{salasnich2002parametric}. Here, similarly to our approach, the potential height is varied periodically to enhance tunnelling. In this case, a Bose-Einstein condensate is discussed, where the time dynamics is governed by the Gross-Pitaevskii equation.

In the future, it will be interesting to study the relation of the SUPER principle to these approaches for parametric resonance in more detail.

\subsection{Practical hints towards realising assisted tunnelling with the Quantum Composer}\label{SUP2}

Assisted tunnelling by swapping between two double-well potentials with different barrier heights $\lambda_1$ and $\lambda_2$ is not an easy task. The parameters have to be chosen carefully to achieve convincing results. We start by choosing two sets of parameters for two different potential functions $V_1(\lambda_1,a,\mu_1)=V_{\rm high}$ and $V_2(\lambda_2,a,\mu_2)=V_{\rm low}$, where $a$ is fixed and would typically be given by the physical system, $\lambda_1 \approx 2\lambda_2$ and $\mu_1=\mu_2$. The closer $\lambda_1$ and $\lambda_2$ are, the closer the two corresponding rotation axes on the Bloch-sphere. Then, the orbits (see Fig. \ref{fig:2Dswingup}) are closer together, resulting in less distance travelled per oscillation period. This translates to more oscillations needed for a full Swing Up. The fidelity graph becomes less steep, and in turn the Swing Up process takes longer. Thus, we choose substantially different $\lambda$.

The Rabi frequencies $\omega_R^{(1)}$ and $\omega_R^{(2)}$ can be found via $E_1^{(1)}-E_2^{(1)}$ and $E_1^{(2)}-E_2^{(2)}$, respectively, that are calculated numerically in Quantum Composer. Using Eq. (\ref{RotA}) and double-angle formulas, we find $\delta/\omega_R=\cos^2{(\theta)}-\sin^2((\theta))=2\cos^2{(\theta)}-1=1-2\sin^2(\theta)$. With (\ref{Tat}), we calculate the initial state $\ket{L_1}$ localised in the left well as 
\begin{align}\label{eq:initialstate}
    \ket{L_1} &= \cos{(\theta)}\ket{E_2} + \sin{(\theta)}\ket{E_1}
    \\
    &= \sqrt{\frac{\omega_R^{(1)}+2\mu a}{2\omega_R^{(1)}}}\ket{E_2}+\sqrt{\frac{\omega_R^{(1)}-2\mu a}{2\omega_R^{(1)}}}\ket{E_1}
\end{align}
using the approximation $\delta \approx 2\mu a$. The desired state can be either chosen as
\begin{equation}\label{eq:R}
    \ket{R_1}=-\sqrt{\frac{\omega_R^{(1)}-2\mu a}{2\omega_R^{(1)}}}\ket{E_2}+\sqrt{\frac{\omega_R^{(1)}+2\mu a}{2\omega_R^{(1)}}}\ket{E_1}
\end{equation}
or
\begin{equation}
    \ket{R_2}=-\sqrt{\frac{\omega_R^{(2)}-2\mu a}{2\omega_R^{(2)}}}\ket{E_2}+\sqrt{\frac{\omega_R^{(2)}+2\mu a}{2\omega_R^{(2)}}}\ket{E_1},
\end{equation}
depending on which of the two potential functions should be final. In our example Swing Up, we chose $\ket{R_2}$.

\begin{figure}
    \centering
    \includegraphics[width=0.95\columnwidth]{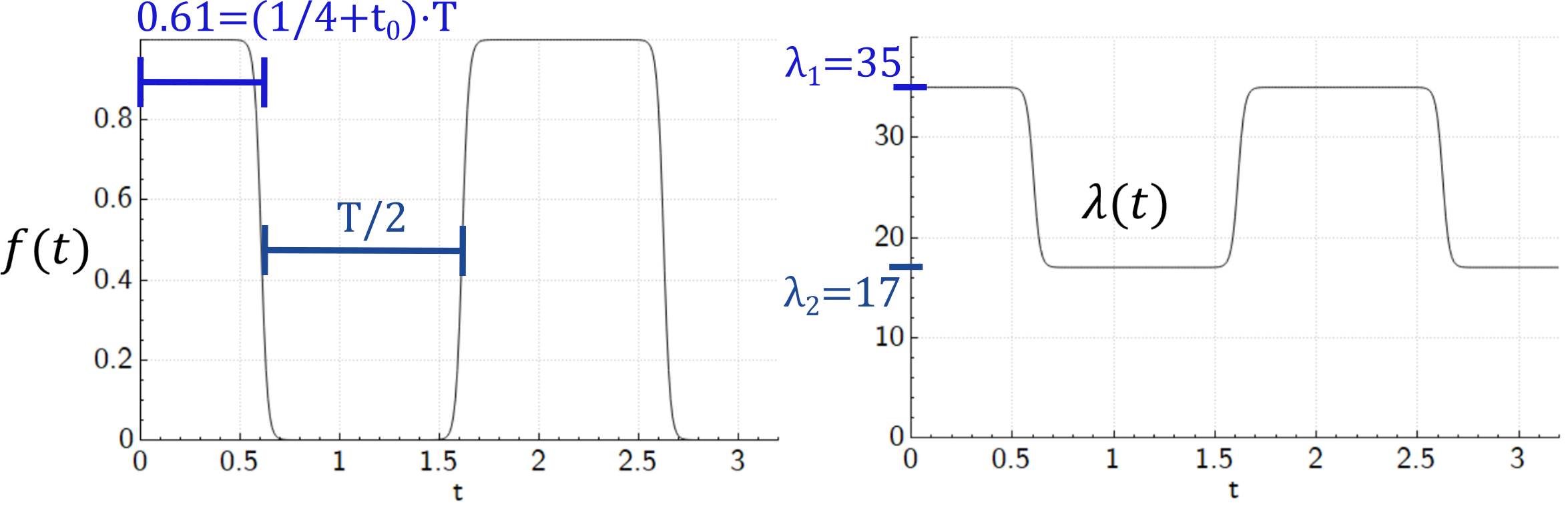}
    \caption{Left: Modulation function $f(t)=\frac{1}{e^{s\cdot g(t)}+1}$ where $s=20$ is the transition slope and $g(t)=-\cos{\left(\frac{t-t_0}{T}\cdot\pi\right)}$ with potential swap time $T/2=1.01$ and time offset $t_0=0.1$. Right: Potential barrier height $\lambda(t)=\lambda_1\cdot f(t)+\lambda_2\cdot(1-f(t))$. The time of the first swap of the potential is $T/4+t_0$.}
    \label{fig:LambdaT}
\end{figure}

Next, the potential swap time $T=\frac{2\pi}{\omega}$( where $\omega$ is the potential oscillation frequency) should be chosen. $T/2$ is the time of one potential being active before swapping to the other. For an optimal fidelity, we observed that $T$ should be about 5\% larger than the average Rabi period $T_R$ of the two potentials, $T_R=4\pi/(\omega_R^{(1)}+\omega_R^{(2)})$. Next, the time offset $t_0$ at the start of the swing-up can be optimised (see Fig. \ref{fig:LambdaT}). $T/4+t_0$ is the time of the first swap of the potential.

\begin{figure}
    \centering
    \includegraphics[width=0.95\columnwidth]{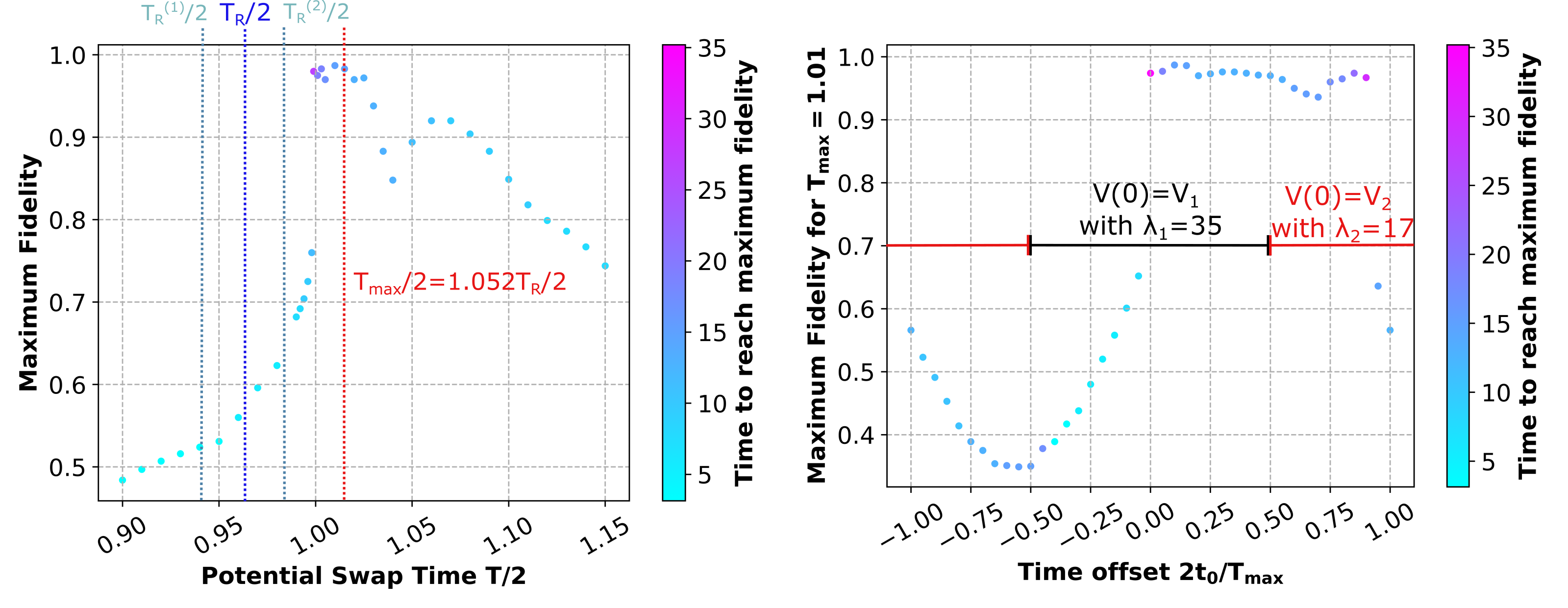}
    \caption{Various maximum fidelities and the times to reach these fidelities for different potential swap times $T/2$ and different time offsets $t_0$ holding the potential swap time constant ($T/4+t_0$ is the time of the first potential swap, see Fig. \ref{fig:LambdaT}). Swapping between two potentials with $\mu=2.8571$, $a=0.7$, $V_1$ with $\lambda_1=35$ and $V_2$ with $\lambda_2=17$. Left: The maximum fidelities reached for various potential swap times $T/2$ at time offsets $t_0=0.1T/2$. Shown are the Rabi oscillation periods of the two potentials $T_{R}^{(1)}/2=\pi/\omega_R^{(1)}$ and $T_{R}^{(2)}/2=\pi/\omega_R^{(2)}$ and the mean Rabi oscillation period $T_R/2$. The maximum fidelity $F=0.987$ is achieved at $T_{\text{max}}/2=1.052 (T_R/2) = 1.01$. This can be further improved to $F=0.994$ by holding the potential constant at $\lambda_2=17$ after $t=15$ as shown in Fig. \ref{fig:assistedTunnelling1}. Right: Maximum fidelities for various different time offsets $2t_0/T_{\text{max}}$ at potential swap time $T_{\text{max}}/2$.
    }
    \label{fig:frequencies}
\end{figure}

Using these parameters, we achieve a high fidelity for assisted tunnelling in the following manner: First, one should optimise the time offset $t_0$ for the starting potential swap time $T$. Next, we optimise $T$ for this offset and repeat. If the resulting fidelity is not satisfying, different values for $\lambda_1, \lambda_2$ or $\mu$ should be chosen. In this manner, a maximum fidelity of over F=0.994 was achieved as is shown in more detail in Fig. \ref{fig:frequencies}.

\begin{figure}
    \centering
    \includegraphics[width=0.6\columnwidth]{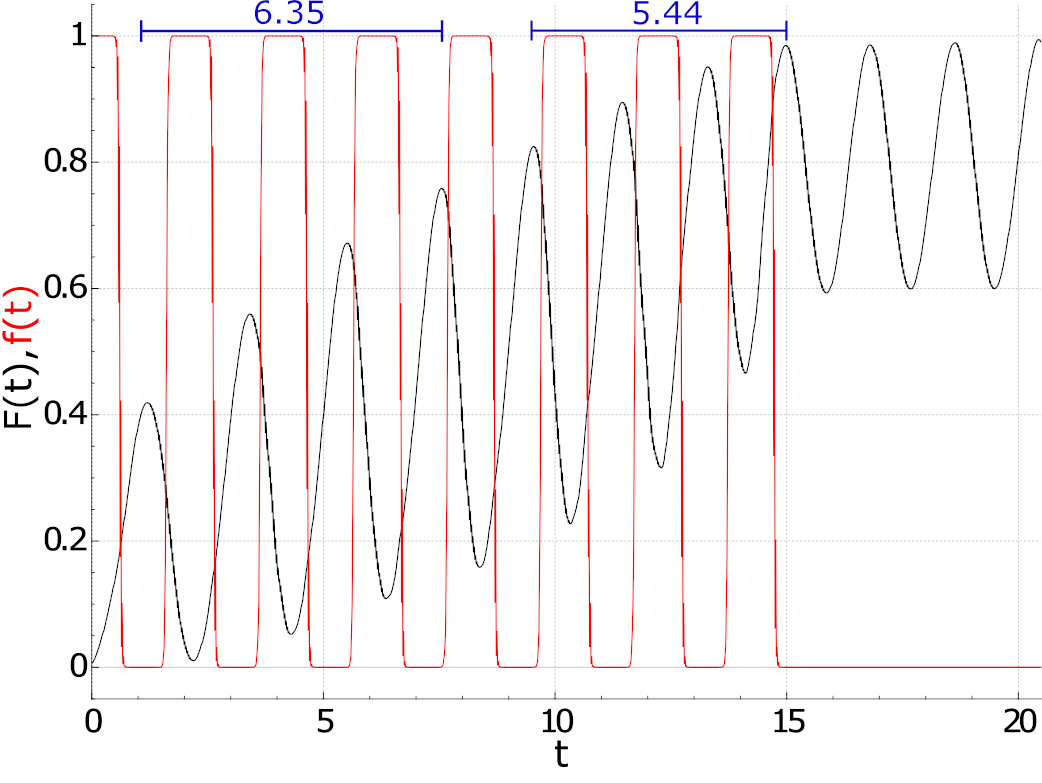}
    \caption{Fidelity F(t) as in Fig. \ref{fig:assistedTunnelling1}, top middle and modulation function $f(t)$ as in Fig. \ref{fig:LambdaT} in red. $f(t)=0$ corresponds to $\lambda=\lambda_1=17$ and $f(t)=1$ corresponds to $\lambda=\lambda_2=35$. The frequency of the Swing Up increases during the process: the first 4 peaks are 6.35 time units apart while the next 4 peaks are only 5.44 time units apart.}
    \label{fig:faster}
\end{figure}

The process is challenging in the sense that it is very sensitive to $T$, $t_0$ and the potential parameters $\lambda$ and $\mu$. This could be because the Rabi frequency is not constant during the Swing Up process. Rather, it increases the further $\ket{\psi(t)}$ travels towards $\ket{R}$ as shown in Fig. \ref{fig:faster}. We have no analytical results that predict this behaviour and can only refer to the numerical results. This can lead to falling out of the swinging rhythm, but can be solved by careful choice of parameters, as shown in this paper.

\subsection{Limits of the Bloch-sphere representation for the double-well system}
\label{lim}

The $2 \times 2$ Hamiltonian (\ref{H0}) includes only the two lowest-lying states of an infinite number of eigenstates in the double-well. In this section, we check the validity of the two-level system approximation (\ref{eq:initialstate}), 
(\ref{eq:R}) for the two lowest-lying states localised in the left and the right well, respectively.

\begin{figure}
    \centering
    \includegraphics[width=0.95\columnwidth]{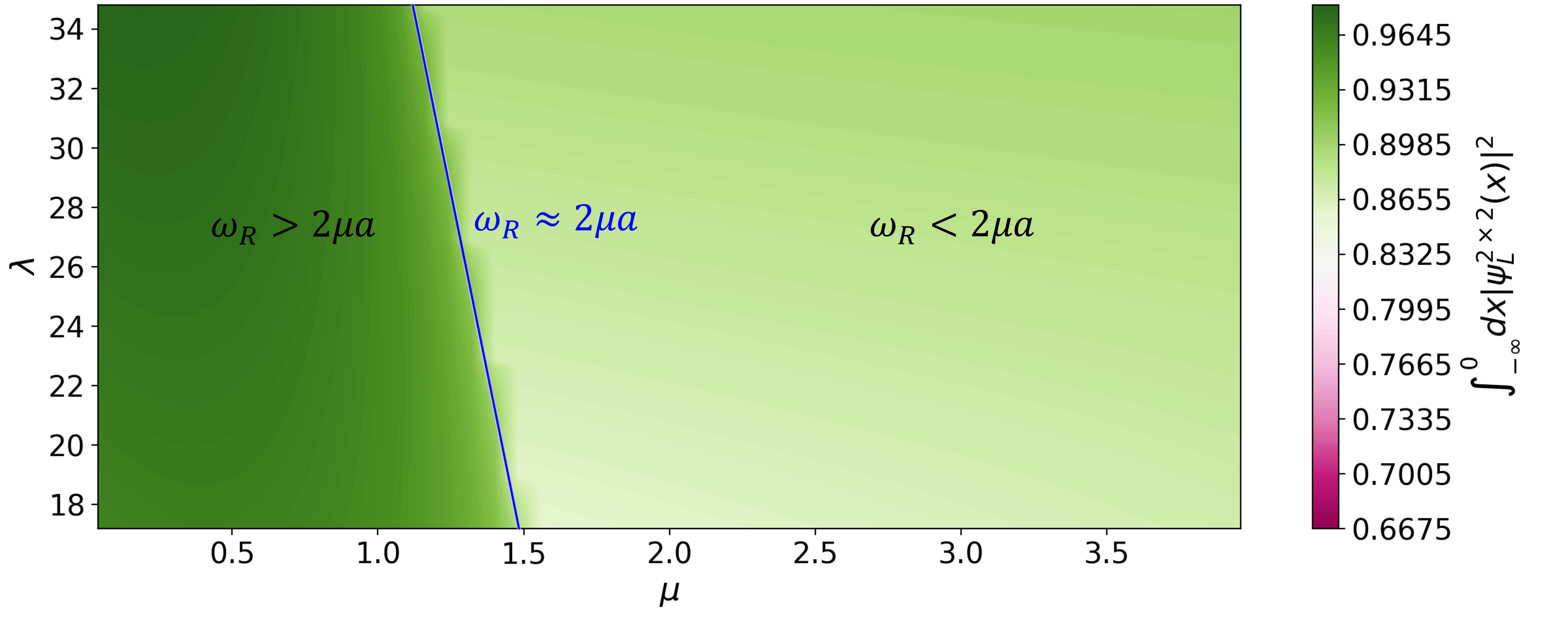}
    \includegraphics[width=0.95\columnwidth]{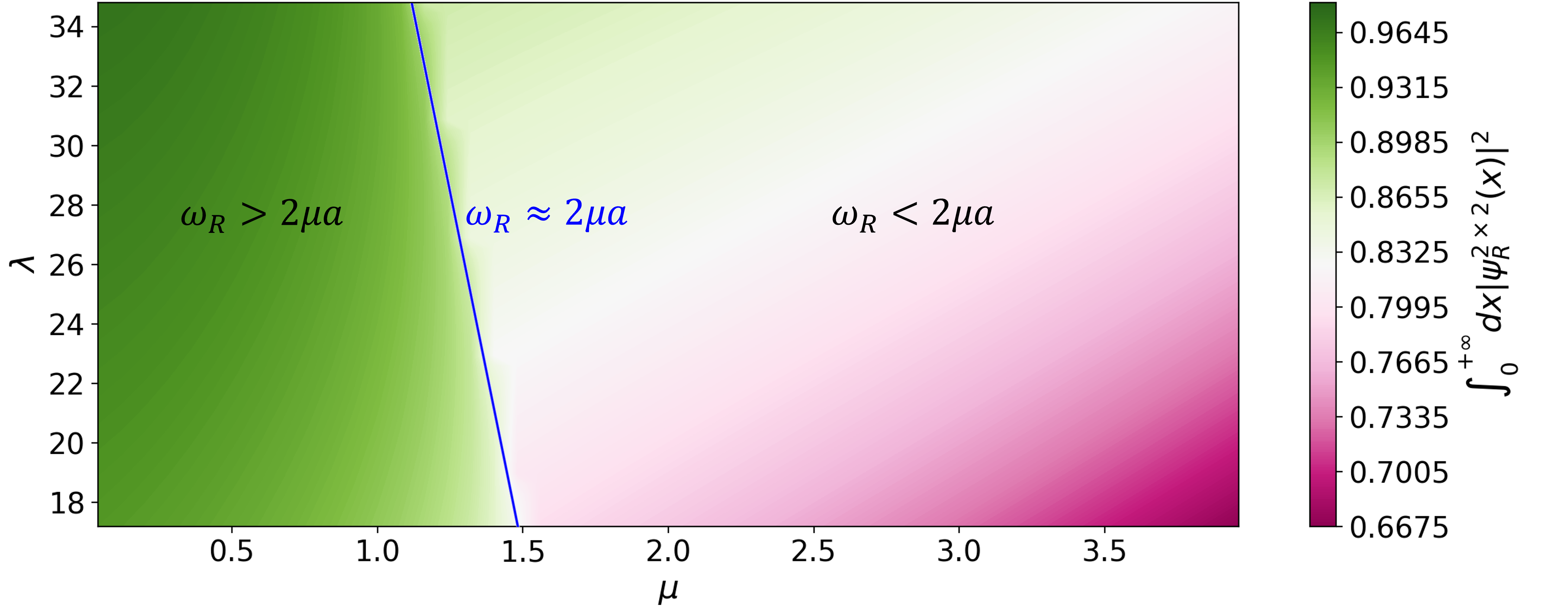}
    \caption{Deviations of the two-level-system approximation by including contributions of higher order levels as defined by (\ref{eer1}), (\ref{eer2}). Top: $\int_{-\infty}^{0}\textrm{d}x|\psi_L^{2 \times 2} (x)|^2$ with $\ket L$ from eq. (\ref{eq:initialstate}) for different $\lambda$ and $\mu$ and $a=0.7$ fixed. Bottom: $\int_{0}^{\infty}\textrm{d}x|\psi_R^{2 \times 2} (x)|^2$ with $\ket R$ from eq. (\ref{eq:R}). In the region $\omega_R>2\mu a$, the deviation from the two-level system is rather small $<10\%$.}
    \label{figL}
\end{figure}

In general, these states are a superposition not only of the first two eigenstates $\ket{E_2}$,$\ket{E_1}$, but include more states $\ket \chi_{R, L}$ with higher energy. Thus, the exact wave functions in position space reads

\begin{eqnarray}
\label{eer1}
      \psi_L^{\rm exact}(x) &=& \bra x L_1 \rangle+\bra x \chi_L \rangle  \equiv 
      \psi_L^{2 \times 2} (x) + \chi_L(x), 
      \\ 
       \psi_R^{\rm exact}(x) &=& \bra x R_1 \rangle+\bra x \chi_R \rangle  \equiv 
       \psi_R^{2 \times 2}(x) + \chi_R(x).
\end{eqnarray}
These states are fully localised either in the left or the right valley, leading to

\begin{eqnarray}
\label{eer2}
      \int_{- \infty}^{0} |\psi_L^{\rm exact}(x)|^2 &=&  \int_{- \infty}^{0}  
      |\psi_L^{2 \times 2} (x)|^2 +  \int_{- \infty}^{0}  |\chi_L(x) |^2 = 1,
      \\ 
      \int_{0}^{\infty}  |\psi_R^{\rm exact}(x)|^2  &=&  \int_{0}^{\infty} |\psi_R^{2 \times 2} (x)|^2 +  \int_{0}^{\infty}  |\chi_R(x) |^2 = 1.
\end{eqnarray}
In Fig. \ref{figL}, we show the expression $1-|\chi_R(x) |^2$ depending on the values for $\lambda, \mu$. In the region with $\omega_R > 2\mu a$, the deviation from the two-level system is below  $5\%$. Note that for $\omega_R = 2\mu a$, the amplitude for $\ket{E_1}$ in (\ref{eq:initialstate}) (and $\ket{E_2}$ in (\ref{eq:R})) vanishes. After the transition to $\omega_R < 2\mu a$, the amplitude becomes imaginary, and the error increases in an asymmetric manner. While the error stays below  $15\%$ for the left valley, it increases to up to $35\%$ in the right valley.

While the visualisation of the SUPER principle on the Bloch sphere is rather intuitive and helps understanding the mechanism, the SUPER principle itself is not restricted to the $2\times2$ level system and indeed includes higher order states during the transition shown in Fig. \ref{fig:assistedTunnelling1}.

\section{Outlook and discussion}
\label{OUT}

Our conclusions can be drawn from a didactical perspective, and from the perspective of recent research in applied physics. While periodic oscillations of potentials have already been studied \cite{GH}, to our knowledge, a systematic enhancement of tunnelling using the SUPER-mechanism has not yet been demonstrated.

In the context of quantum tunnelling, it seems possible to find experimental realisations for a "high" value and a "low" value of the potential barrier. Explicitly, we may think of a Bose-Einstein condensate trapped in a double-well potential. An experimental realisation, where the atoms are magnetically coupled to a single-mode of the microwave field inside a superconducting resonator, has been reported in \cite{Chu2011}. By modulating the magnetic field strength, the trap potential could be manipulated to embark for assisted tunnelling.

As another model system, we may think of a double-well potential realised for an individual fluxon trapped in a Josephson tunnel junction. Again, parameters of the potential can be manipulated by the strength of a magnetic field \cite{Monaco}. Since the double-well potential is the smallest subpart of the general class of Hubbard-model systems in condensed matter physics, many other applications of assisted tunnelling might be of interest.

Our main concern however, is physics education: Since the Quantum Composer is available for free, it is possible to discuss Rabi oscillations in an intuitive and direct manner using this powerful modelling tool on almost any computer system. The result of the educator-developer dialogue is the updated version of the Composer. With that, the explicit visualisation of the Bloch sphere dynamics is possible. Using the Bloch-sphere representation, the SUPER principle has a simple and general geometric interpretation as shown in Figs. \ref{fig:Blochspheres2} and \ref{fig:2Dswingup} and even a knot theoretic interpretation \cite{Heu2019b}. Translated back to the double-well, we may apply this mechanism to realise assisted tunnelling by introducing an appropriate oscillation of the potential height.

Explicitly, at high school, the dependence of the Rabi frequency on the potential parameters $(\lambda, a, \mu)$ can qualitatively be discussed using just the graphical interface provided by the Quantum Composer. On the undergraduate level, the quantitative relations explored in Sect. \ref{TDW} can be accessed, in particular, the dependence of $S_{\rm inst.} \propto \sqrt{\lambda} a^3$ on the level spacing of the two lowest eigenvalues, $(E_1 - E_2) \propto \exp{(-S_{\rm inst.})}$. On the graduate level, assisted tunnelling and the SUPER principle can be explored numerically as shown in Sect. \ref{SUP}. Here, numeric results can be used to provide a qualitative understanding of the SUPER-mechanism as shown in Fig. \ref{fig:2Dswingup}. The multiple representation that the Quantum Composer offers, like the state evolution plot and the Bloch sphere, can enhance understanding of this concept. All Quantum Composer files we used in our work can be downloaded from \cite{Animations}.

Concerning research in physics, still many questions remain open. It would be desirable to understand the relation between the numerical results from the Quantum Composer and analytic results, such as \ref{coup}, even better. Moreover, analytic results for assisted tunnelling are not yet available, and it would be desirable to analytically explore this approach further in accordance with \cite{Zhi2021}.

An empirical study on solving tasks with multiple representations in quantum composer has been conducted recently in an eye-tracking study \cite{küchemann2023impact}. Similarly, a possible direction of further didactical research is the exploration of the benefits and challenges of teaching approaches using visualisation tools such as the Quantum Composer, with various target groups and in different contexts like the one described in this paper.

\section*{Acknowledgements}

The Aarhus team acknowledged funding from Carlsberg Foundation, Independent Research Fund Denmark (ReGAME) and European Research Council (PQTEI).

\noindent The M\"unster team is grateful to D. Reiter and T. Bracht for valuable discussions.

\noindent J.B. acknowledges support by the project QuanTUK at the RPTU in Kaiserslautern, supported by the Federal Ministry of Education and Research (FKZ13N15995).

\section*{References}
\bibliography{main}

\begin{thebibliography}{10}

\bibitem{Sakurai}
J.~Sakurai.
\newblock {\em Modern quantum mechanics, 2nd edition}.
\newblock {Cambridge University Press}, 2017.

\bibitem{Gilbert}
J.~Gilbert, M.~Reiner, and M.~Nakhleh.
\newblock {\em Visualization: Theory and Practice in Science Education}.
\newblock Cambridge University Press, 2008.

\bibitem{R2}
A.~Ahmed, J.~J. Jensen, C.~A. Weidner, J.~J. Sørensen, M.~Mudrich, and J.~F. Sherson.
\newblock Quantum composer: A programmable quantum visualization and simulation tool for education and research.
\newblock {\em Arxiv, physics.ed-ph}, 2020.

\bibitem{Jelic_2012}
V~Jelic and F~Marsiglio.
\newblock The double-well potential in quantum mechanics: a simple, numerically exact formulation.
\newblock {\em European Journal of Physics}, 33(6):1651, sep 2012.

\bibitem{Song_2008}
Dae-Yup Song.
\newblock Tunneling and energy splitting in an asymmetric double-well potential.
\newblock {\em Annals of Physics}, 323(12):2991--2999, dec 2008.

\bibitem{GH}
M.~Grifoni and P.~Hänggi.
\newblock Driven quantum tunneling.
\newblock {\em Phys. Rep.}, 304(229), 1998.

\bibitem{PhysRevLett.98.263601}
G.~Della~Valle, M.~Ornigotti, E.~Cianci, V.~Foglietti, P.~Laporta, and S.~Longhi.
\newblock Visualization of coherent destruction of tunneling in an optical double well system.
\newblock {\em Phys. Rev. Lett.}, 98:263601, Jun 2007.

\bibitem{PhysRevLett.100.190405}
E.~Kierig, U.~Schnorrberger, A.~Schietinger, J.~Tomkovic, and M.~K. Oberthaler.
\newblock Single-particle tunneling in strongly driven double-well potentials.
\newblock {\em Phys. Rev. Lett.}, 100:190405, May 2008.

\bibitem{Bukov_2014}
Marin Bukov, Luca Dalessio, and A.~Polkovnikov.
\newblock Universal high-frequency behavior of periodically driven systems: from dynamical stabilization to floquet engineering.
\newblock {\em Advances in Physics}, 64, 07 2014.

\bibitem{PhysRev.51.652}
I.~I. Rabi.
\newblock Space quantization in a gyrating magnetic field.
\newblock {\em Phys. Rev.}, 51:652--654, Apr 1937.

\bibitem{Rei2021}
T.~Bracht, M.~Cosacchi, T.~Seidelmann, M.~Cygorek, A.~Vagov, M.~Axt, T.~Heindel, and D.~Reiter.
\newblock Swing-up of quantum emitter population using detuned pulses.
\newblock {\em PRX QUANTUM}, 2(040354), 2021.

\bibitem{Rei2022}
K.~Yusuf, F.~Kappe, V.~Remesh, T.~Bracht, J.~Muenzberg, S.~Covre~da Silva, T.~Seidelmann, V.~Axt, A.~Rastelli, D.~Reiter, and G.~Weihs.
\newblock Super scheme in action: Experimental demonstration of red-detuned excitation of a quantum dot.
\newblock {\em https://arxiv.org/abs/2203.00712}, 2022.

\bibitem{quatomic}
Carrie~A. Weidner, Shaeema~Z. Ahmed, Jesper H.~M. Jensen, and Jacob~F. Sherson.
\newblock Publications using quatomic software.
\newblock \url{https://www.quatomic.com/composer/}.
\newblock Retrieved 3/25/2021.

\bibitem{kohnle2012new}
Antje Kohnle, Donatella Cassettari, Tom~J Edwards, Callum Ferguson, Alastair~D Gillies, Christopher~A Hooley, Natalia Korolkova, Joseph Llama, and Bruce~D Sinclair.
\newblock A new multimedia resource for teaching quantum mechanics concepts.
\newblock {\em American Journal of Physics}, 80(2):148--153, 2012.

\bibitem{mckagan2008developing}
SB~McKagan, Katherine~K Perkins, Michael Dubson, Chris Malley, Sam Reid, Ron LeMaster, and CE~Wieman.
\newblock Developing and researching phet simulations for teaching quantum mechanics.
\newblock {\em American Journal of Physics}, 76(4):406--417, 2008.

\bibitem{uni-heidelberg}
Mctdhb-lab: The laboratory to study quantum many-body dynamics.
\newblock \url{https://www.pci.uni-heidelberg.de/tc/usr/mctdhb/MCTDHB-Lab.html} (accessed 14th June 2023).

\bibitem{sorensen2019qengine}
J~Jensen S{\o}rensen, JHM Jensen, T~Heinzel, and Jacob~Friis Sherson.
\newblock Qengine: A c++ library for quantum optimal control of ultracold atoms.
\newblock {\em Computer Physics Communications}, 243:135--150, 2019.

\bibitem{R1}
Carrie~A. Weidner, Shaeema~Z. Ahmed, Jesper H.~M. Jensen, and Jacob~F. Sherson.
\newblock Research using quatomic software.
\newblock \url{https://www.quatomic.com/composer/research}.
\newblock Retrieved 27/11/2021.

\bibitem{k}
J~Jensen, J.~J Sørensen, K.~Mølmer, and J.~F. Sherson.
\newblock Time-optimal control of collisional $\sqrt{\mathrm{swap}}$ gates in ultracold atomic systems.
\newblock {\em Phys. Rev. A}, 100:052314, 2019.

\bibitem{a}
R.~Bell and H~Bianchi.
\newblock The many levels of inquiry.
\newblock {\em Science \& Children}, 46(2), 2008.

\bibitem{b}
L.~V Rodriguez, J.~T. van~der Veen, A.~Anjewierden, E.~van~den Berg, and T.~de~Jong.
\newblock Designing inquiry-based learning environments for quantum physics education in secondary schools.
\newblock {\em Phys. Educ.}, 55(6), 2020.

\bibitem{c}
I.~Testa, A.~Colantonio, S.~Galano, I.~Marzoli, F.~Trani, and U.~Scotti~di Uccio.
\newblock Effects of instruction on students’ overconfidence in introductory quantum mechanics.
\newblock {\em Phys. Rev. Phys. Educ. Res.,}, 16(1):010143, 2020.

\bibitem{d}
Z.~C. et~al. Seskir.
\newblock Quantum games and interactive tools for quantum technologies outreach and education.
\newblock {\em Opt. Eng.}, 61(8):081809, 2022.

\bibitem{e}
S.~Goorney, L.~Foti, C.and~Santi, J.~Sherson, J.~Yago~Malo, and M.L Chiofalo.
\newblock Culturo-scientific storytelling.
\newblock {\em Education Sciences}, 12(7):474, 2022.

\bibitem{g}
C.~Singh.
\newblock Interactive learning tutorials on quantum mechanics.
\newblock {\em American Journal of Physics}, 76(4):400--405, 2008.

\bibitem{h}
D.~Kuhn.
\newblock {\em What is Scientific Thinking and How Does it Develop?, in The Wiley-Blackwell Handbook of Childhood Cognitive Development}.
\newblock {Blackwell Publishing Ltd}, 2008.

\bibitem{Z}
A.~Ahmed, C.~A. Weidner, J.~J Jensen, J.~F. Sherson, and H.~J. Lewandowski.
\newblock Student use of a quantum simulation and visualization tool.
\newblock {\em Eur. J. Phys.}, 43(6), 2022.

\bibitem{i}
O~Elíasson, J.~S. Laustsen, R.~Heck, R.~Müller, J.~J. Arlt, C.~A Weidner, and J.~F. Sherson.
\newblock Spatial tomography of individual atoms in a quantum gas microscope.
\newblock {\em Phys. Rev. A}, 102:053311, Nov 2020.

\bibitem{Animations}
Modeling assisted tunneling with the quantum composer using the super-principle.
\newblock \url{https://physikkommunizieren.de/quantum/tunneling/}, August 2023.

\bibitem{hopf_fib}
Ira~S. Moskowitz.
\newblock The bloch sphere for topologists.
\newblock {\em NAVAL RESEARCH LAB WASHINGTON DC CENTER FOR HIGH ASSURANCE COMPUTING SYSTEMS (CHACS)}, 05 2008.

\bibitem{Miao}
G.~Miao.
\newblock Reports on progress in physics tunneling path toward spintronics.
\newblock {\em Rep. Prog. Phys.}, 74(036501), 2011.

\bibitem{AB}
D.~Ananikian and T.~Bergeman.
\newblock Gross-pitaevskii equation for bose particles in a double-well potential: Two-mode models and beyond.
\newblock {\em Physical Review A}, 73(013604), 2006.

\bibitem{DWM}
Michel~H. Devoret, Andreas Wallraff, and John~M. Martinis.
\newblock Superconducting qubits: A short review.
\newblock {\em arXiv: Mesoscale and Nanoscale Physics}, 2004.

\bibitem{Stade}
K.~Stademann.
\newblock Analysis of secondary school quantum physics curricula of 15 different countries: Different perspectives on a challenging topic.
\newblock {\em PHYSICAL REVIEW PHYSICS EDUCATION RESEARCH}, 15(010130), 2019.

\bibitem{Rast}
G.~Rastelli.
\newblock Semi-classical formula for quantum tunneling in asymmetric double-well potentials.
\newblock {\em Phys RevA.}, 86(01210), 2012.

\bibitem{Cole2}
S.~Coleman.
\newblock {\em Aspects of Symmetry: Selected Erice Lectures}.
\newblock {Cambridge Univ. Press}, 1985.

\bibitem{Peano}
V.~Peano, M.~Marthaler, and M.~I. Dykman.
\newblock Sharp tunneling peaks in a parametric oscillator: Quantum resonances missing in the rotating wave approximation.
\newblock {\em Phys. Rev. Lett.}, 109:090401, Aug 2012.

\bibitem{Golo_2005}
V.~L. Golo and Yu.~S. Volkov.
\newblock Chaotic tunneling in a laser field.
\newblock {\em Journal of Experimental and Theoretical Physics Letters}, 82(4):181--184, aug 2005.

\bibitem{Salasnich_2002}
L~Salasnich, A~Parola, and L~Reatto.
\newblock Periodic quantum tunnelling and parametric resonance with cigar-shaped bose-einstein condensates.
\newblock {\em Journal of Physics B: Atomic, Molecular and Optical Physics}, 35(14):3205--3216, jul 2002.

\bibitem{salasnich2002parametric}
Luca Salasnich.
\newblock Parametric resonance phenomena in bose-einstein condensates: Enhanced quantum tunneling of coherent matter pulses, 2002.

\bibitem{Chu2011}
H.~Ng and S.~Chu.
\newblock Steady-state entanglement in a double-well bose-einstein condensate through coupling to a superconducting resonator.
\newblock {\em Phys. Rev. A}, 84(023629), 2011.

\bibitem{Monaco}
R.~Monaco.
\newblock Engineering double-well potentials with variable-width annular josephson tunnel junctions.
\newblock {\em Journal of Physics: Condensed Matter}, 28(445702), 2016.

\bibitem{Heu2019b}
S.~Heusler and M.~Ubben.
\newblock A haptic model of entanglement, gauge symmetries and minimal interaction based on knot theory.
\newblock {\em Symmetry}, 11(11):1399, 2019.

\bibitem{Zhi2021}
Z.~Shi, Y.~Chen, W.~Qin, Y.~Xia, X.~Yi, S.~Zheng, and F.~Nori.
\newblock Two-level systems with periodic n-step driving fields: Exact dynamics and quantum state manipulations.
\newblock {\em PHYSICAL REVIEW A}, 104(053101), 2021.

\bibitem{küchemann2023impact}
Stefan Küchemann, Malte Ubben, David Dzsotjan, Sergey Mukhametov, Carrie~A. Weidner, Linda Qerimi, Jochen Kuhn, Stefan Heusler, and Jacob~F. Sherson.
\newblock The impact of an interactive visualization and simulation tool on learning quantum physics: Results of an eye-tracking study, 2023.

\end{thebibliography}
\bibliographystyle{unsrt}

\end{document}